\documentclass[aps,prl,twocolumn,10pt,notitlepage,superscriptaddress,nofootinbib,showkeys,letterpaper]{revtex4-1}
\usepackage{amstext,amsmath,amssymb,amsfonts,bbm, amsthm}

\usepackage[latin1]{inputenc}

\usepackage{fancyhdr}
\usepackage{graphicx}
\usepackage{hyperref}

\usepackage{epstopdf}
\usepackage{rotating}
\usepackage{cleveref}

\usepackage{float}

\usepackage{cases}
\usepackage{tabls}

\usepackage{subfigure}

\usepackage{amsthm}

\usepackage{tocvsec2}

\usepackage{color}

\def\beq{\begin{equation}}
\def\be{\begin{equation}}
\def\ee{\end{equation}}
\def\bes{\begin{eqnarray}}
\def\ees{\end{eqnarray}}

\def\sims{\underset{t\to\infty}{\sim}}

\begin{document}

\title{\large \bf Universal Statistics of Selected Values}

\author{Matteo Smerlak}\email{msmerlak@perimeterinstitute.ca}
\affiliation{Perimeter Institute for Theoretical Physics, 31 Caroline St.~N., Waterloo ON N2L 2Y5, Canada}
\author{Ahmed Youssef}\email{youssef@ld-research.com}
\affiliation{LD - Research, Pappelallee 78/79, 10437 Berlin, Germany}

\date{\small\today}

\begin{abstract}
Selection, the tendency of some traits to become more frequent than others in a population under the influence of some (natural or artificial) agency, is a key component of Darwinian evolution and countless other natural and social phenomena. Yet a general theory of selection, analogous to the Fisher-Tippett-Gnedenko theory of extreme events, is lacking. Here we introduce a probabilistic definition of selection and show that selected values are attracted to a universal family of limiting distributions. The universality classes and scaling exponents are determined by the tail thickness of the random variable under selection. Our results are supported by data from molecular biology, agriculture and sport. 
\end{abstract} 

\maketitle

\paragraph{\textbf{Introduction}.} In a posthumous manuscript \cite{Price:1995hsa},\footnote{Price apparently wrote this manuscript after publishing his famous covariance equation \cite{Price:1970ez}, indicating that he was aware of its limitations (which are the same as the limitations of the Fisher fundamental theorem discussed below). Yet the term ``selection theory" is often incorrectly identified with the Price equation in the literature.} 
the population geneticist G. Price noted that ``selection has been studied mainly in genetics, but of course there is much more to selection than just genetical selection". He gave  examples of selection processes relevant to psychology, chemistry, archeology, linguistics, history, economics and epistemology, and remarked that ``despite the pervading importance of selection in science and life, there has been no abstraction and generalization from genetical selection to obtain a general selection theory."  

Price stressed two key features of the theory to be developed. First, selection should be studied as a mathematical transformation, irrespective of the (natural or artificial) agency responsible for that transformation. Second, selection theory should encompass both ``subset selection", wherein a subset is picked out from a set according to some criterion, and ``Darwinian selection", dominance through differential reproduction. If such a general concept could be formulated mathematically, he thought, it would have an impact comparable to Shannon's formal theory of communication \cite{Shannon1948}. 

Whether or not the analogy is apt, there is a clear need for a general theory of selection. In biology, identifying signatures of natural selection (in particular at the genotypic level \cite{Nielsen:2005kx}) is a fundamental problem with important applications, for instance in the context of cancer research \cite{Huang:2012kk}. Such a theory would also be useful for the development of selection-based search methods throughout the sciences, including genetic algorithms \cite{Galletly:2013bu} in computer science or SELEX protocols \cite{Bouchard:2010gn} in pharmacology. It would also provide a conceptual framework for the current widespread interest in analytics in sport, education, academia and other competitive fields where selection plays a key role. In spite of a handful of formal explorations \cite{Gorban:2007il,Grafen:2006dq,Karev:2010bc}---and forty-five years after Price's comments---selection theory is still ``a theory waiting to be born" \cite{Price:1995hsa}.\footnote{Somewhat paradoxically, rather sophisticated evolutionary models involving selection \textit{and} mutations, drift, gene flow, etc. are well developed \cite{Tsimring:1996cra,Brunet:1997hxa,Fisher:2013ck}.}

The fundamental question selection theory should address was clearly articulated in a recent paper by Boyer \textit{et al.} on molecular evolution \cite{Boyer:2015us}. The authors considered large libraries of randomized biomolecules which, in the spirit of SELEX, they selected on the basis of their affinity for a molecular target of interest. As they noted, ``merely counting the number of different individuals provides a poor indication of the potential of a population to satisfy a new selective constraint". The key problem, then, is how to identify the features of the population which characterize its selective potential. How diverse should it be? How should we measure this ``diversity"? And how does a population with a given selective potential respond to selection pressures of different strengths?

In this paper we explore some of the most basic statistical aspects of the selection process. To this effect we define a \textit{selected value} as the transformation $S_tW$ of a non-negative random variable $W$ given by \begin{equation}\label{sel}
\mathbb{P}(S_tW=w)\propto w^t\mathbb{P}(W=w),
\end{equation}
for some parameter $t>0$. This definition is in the spirit of the one proposed by Price\footnote{Price wrote ``Selection on a set in relation to property $W$ is the act or process of producing a corresponding set in a way such that the amounts of each entity are non-randomly related to the corresponding $W$ values" \cite{Price:1995hs}, which we can write $\mathbb{P}(SW=w)\propto f(w)\mathbb{P}(W=w)$ for an arbitrary function $f(w)$. If this function is one-to-one and monotone we can reduce it to $f(w)=w$ by a suitable change of variable.}, and furthermore it has the advantage of carrying a natural semi-group structure ($S_{t'}\circ S_t=S_{t+t'}$) from which notions of ``weak selection" ($t\to 0$) and ``strong selection" ($t\to\infty$) can be defined. Moreover \eqref{sel} has a very intuitive Darwinian interpretation: if $W$ represents the number of viable offspring of an organism in a heterogenous population (its evolutionary ``fitness"), then $S_tW$ describes the change in the distribution of fitness after $t$ generations. Note,  however,  that \eqref{sel} is equally consistent with subset selection: the variable $S_tW$ may represent a subset of a population biased towards larger values of $W$, in such a way that an entity with $W=2w_0$ is $2^t$ more likely to be picked than an entity with $W=w_0$. 

%
%
%

\begin{figure}
	\includegraphics[width=.45\textwidth]{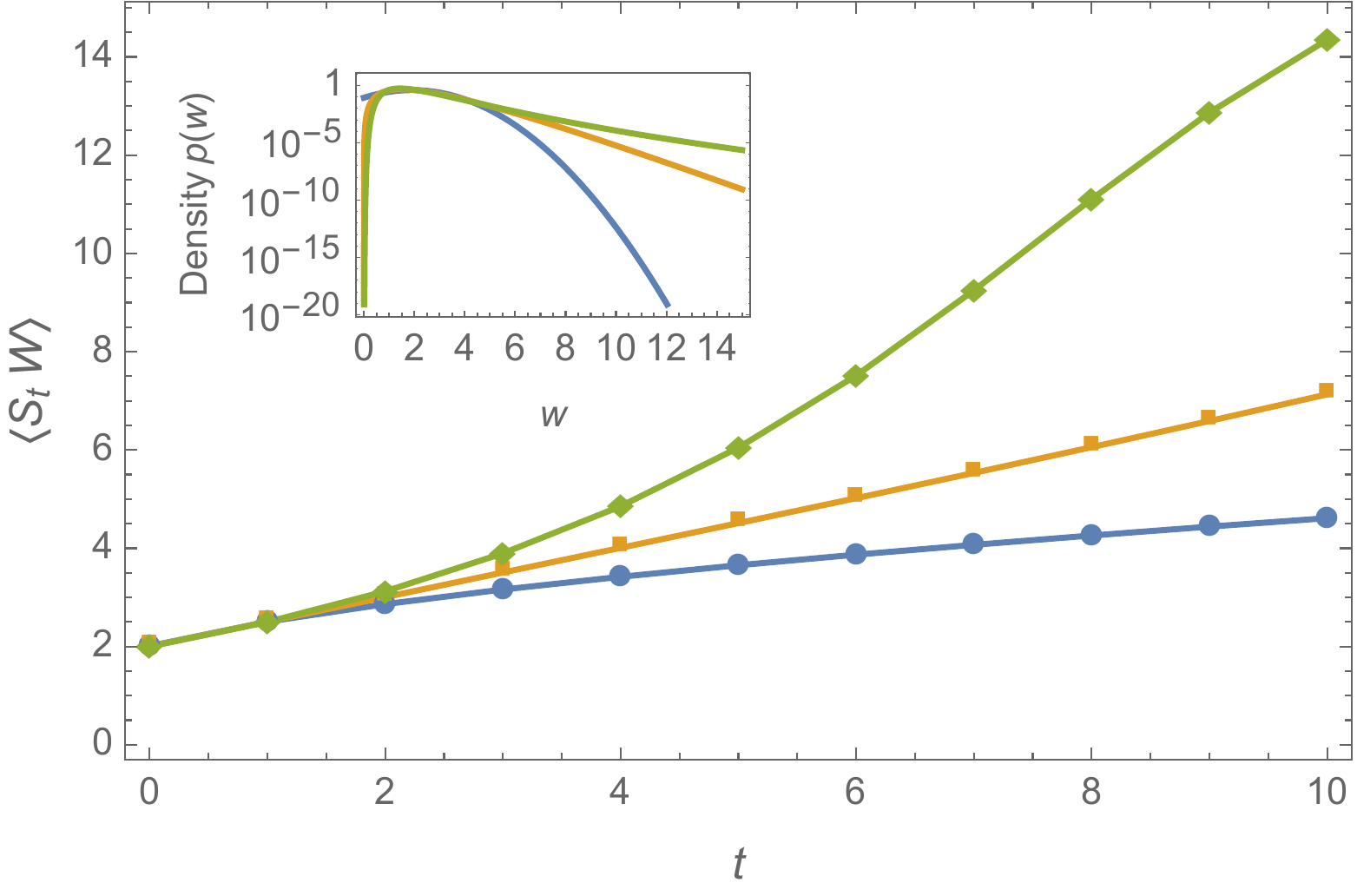}
	\caption{The mean and variance of the ancestral value $W$ are often used to characterize the selective potential of a population, but they do not accurately predict mean-fitness trajectories in the long run. In this simulation we started from populations of $10^6$ values sampled from three distribution with identical mean 2 and variance 1: a normal distribution (blue), a gamma distribution (orange) and a log-normal distribution (green). This distributions have markedly different tail behavior (inset), leading to sub-linear, linear and super-linear growth of mean fitness respectively. At late times finite size effects dominate.}
	\label{varianceno}
\end{figure}

\paragraph{\textbf{Fisher's fundamental theorem}.} The best known result concerning the relation between the selective potential of a population and its diversity is Fisher's ``fundamental theorem of natural selection" \cite{Fisher:1930wya}. In the language of evolutionary theory, Fisher's theorem states that the rate of growth of a population mean fitness under selection is proportional to its variance in fitness. In our notations this reads
\begin{equation}\label{fisher}
	\langle S_{t+1}W\rangle-\langle S_{t}W\rangle=\frac{\textrm{var}(S_{t}W)}{\langle S_{t}W\rangle}.
\end{equation}
This identity---an easy consequence of \eqref{sel}---captures a basic aspect of  selection dynamics: the larger the variation in fitness at a given time, the faster evolution proceeds, or ``variation is the fuel of evolution" as the catchphrase goes. In the limit where all lineages have the same fitness, $\textrm{var}(S_tW)=0$, the mean fitness stops growing and evolution comes to a halt. (This does not imply that all individuals have become identical, or even that they all reproduce at the same rate: all that matters is that they all have the same number of descendants.) Fisher was impressed by the generality of Eq. \eqref{fisher} and compared it to the second law of thermodynamics \cite{Fisher:1930wya}. Later it was realized that various complications (such as mutations, frequency dependence or finite size effects) limit the relevance of Fisher's theorem for biological evolution \cite{Orr:2009kh}. More importantly for our purpose, \eqref{fisher} does not predict the behavior of $\langle S_tW\rangle$ as a function of $t$ and $W$, a shortcoming sometimes referred to as ``dynamic insufficiency"  \cite{Lewontin:1974tea,Barton:1987hla,Frank:1997ws}. Fig. \ref{varianceno} plots $\langle S_tW\rangle$ for three different ancestral distributions with equal mean and variance: the divergence of the trajectories illustrates that neither $\langle W\rangle$ nor $\textrm{var}(W)$ are good predictors of $S_tW$ beyond the short-term or weak selection regime $t\simeq 0$. To make progress, a different approach is needed. As we now show, the key is to focus not on the moments of $W$, but rather on its tail structure.\footnote{This is was hinted at in \cite{Boyer:2015us}, but our specific conclusions are different.}
%
%
%

\paragraph{\textbf{Assumptions and further definitions}.} We assume that the variable $W$ has an absolutely continuous density $p(w)$, i.e. we exclude discrete variables and small population sizes. Next we distinguish two cases: 
\begin{itemize}
	\item \textit{Positive selection}. The variable $W$ has unbounded support $\Sigma$, viz. $\sup\Sigma=\infty$. 
	\item \textit{Negative selection}. The variable $W$ has a finite right end-point $\sup\Sigma\equiv w_+<\infty$. 
\end{itemize}
These two cases are idealizations: in practice, positive selection occurs when $\langle W\rangle \ll w_+$, while negative selection corresponds to $\langle W\rangle \approx w_+$.  In evolutionary terms we can think of these idealizations as capturing respectively the dynamics of rapid adaptation and of evolutionary stasis. Crossovers between these two regimes are possible, as explained below. 

\begin{figure*}
	\includegraphics[width=.47\textwidth]{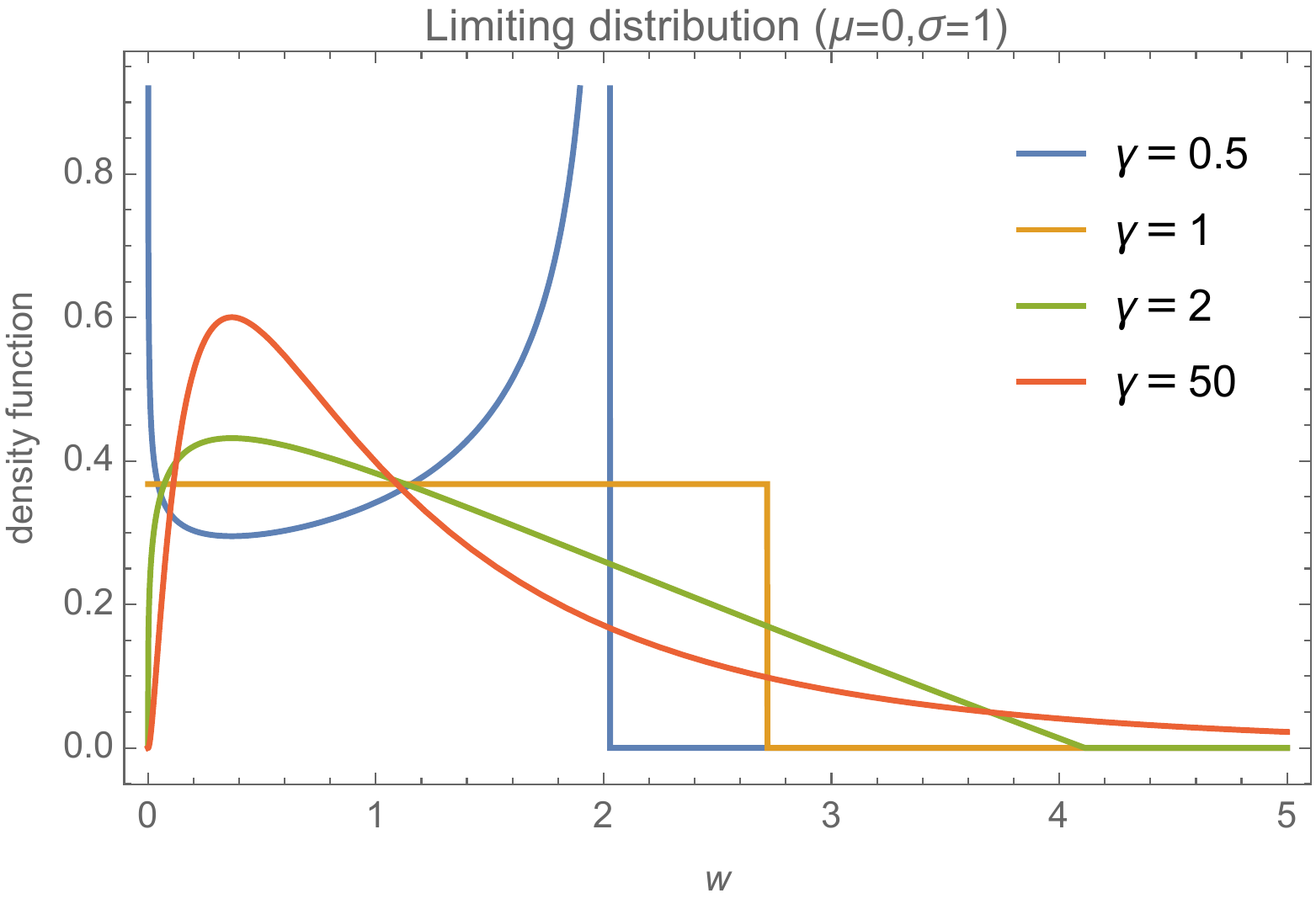}\hfill\includegraphics[width=.47\textwidth]{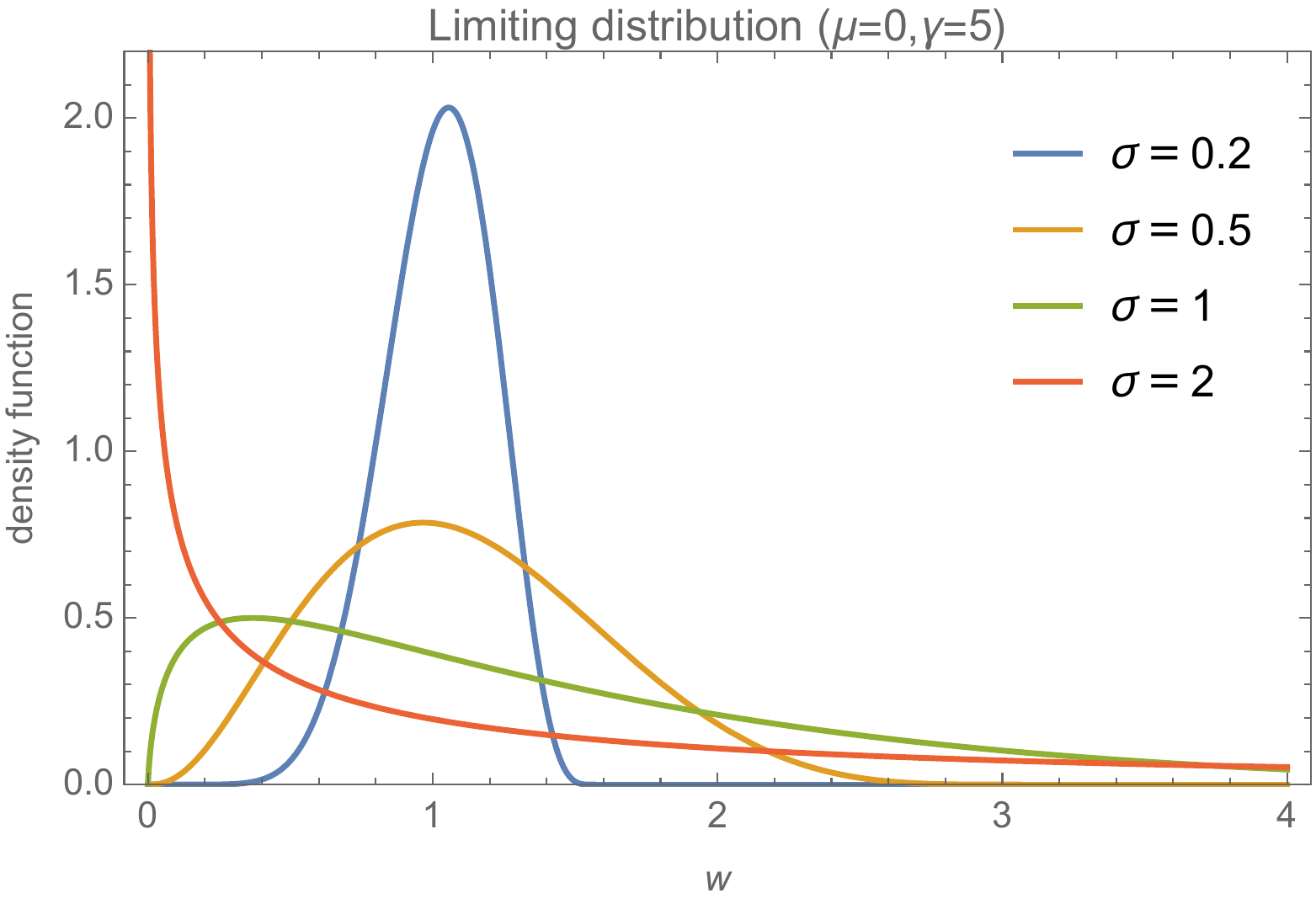}
	\caption{The limiting density functions \eqref{density} for several values of $\gamma$ (left) and $\sigma$ (right).}
	\label{densityFig}
\end{figure*}

Third, we characterize the tail behavior of fitness distributions. To that effect we consider the tail function
\begin{equation}
	T(w)\equiv\int_w^{\sup\Sigma}p(w')dw',
\end{equation}
giving the fraction of individuals with fitness at least  $w$. $T(w)$ goes to zero when $w$ approches $\sup\Sigma$ with a rate that measures the thickness of the tail of $W$. How exactly this rate should be defined requires some further distinctions:
\begin{itemize}
	\item \textit{Positive selection}. For unbounded variables we distinguish between light and heavy tails. We say that $W$ has a \textit{light tail} with index $\alpha$ if\footnote{This condition can be generalized in terms of the notion of regularly varying function \cite{BINGHAM:1989ca}.}
		\begin{equation}\label{defalpha}
		-\ln T(w)\underset{w\to\infty}{\sim}Aw^\alpha\quad\textrm{for some}\ A>0
	\end{equation}
	and a \textit{heavy tail} with index $\beta$ if 
	\begin{equation}\label{defbeta}
		-\ln T(w)\underset{w\to\infty}{\sim}B(\ln w)^\beta\quad\textrm{for some}\ B>0.
	\end{equation}
	In either case we define the location $\mu$ and scale $\sigma$ of $W$ by $\mu=\langle\ln W\rangle$ and $\sigma^2=\langle (\ln W-\mu)^2\rangle$. 
	\item \textit{Negative selection}. We say that a variable with finite right end-point $w_+$ has a short tail with index $\gamma>0$ if
	\begin{equation}\label{defgamma}
		T(w)\underset{w\to w_+}{\sim}C(w_+-w)^{\gamma}\quad\textrm{for some}\ C>0.
		\end{equation}
\end{itemize}		
Note that not every distribution satisfies these assumptions. Power-law distributions, in particular, have unbounded support but do not fall in the classes \eqref{defalpha} and \eqref{defbeta}. We exclude them because they blow up at finite $t$ under the selection dynamic \eqref{sel}. 

%
%
\paragraph{\textbf{Limiting distributions.}}

To analyze the behavior of $S_tW$ when $t$ becomes large we proceed in three steps. First, we pass to  $\ln S_tW$ and consider the associated density function $\pi_t(x)$. Second, we consider the cumulant-generating function of $\pi_t(x)$, defined by $\psi_t(\nu)=\ln\int e^{\nu x}\pi_t(x)dx$. In terms of $\psi_t(\nu)$ the selection equation \eqref{sel} reads 
\begin{equation}\label{cgfflow}
	\psi_t(\nu)=\psi(\nu+t)-\psi(t),
\end{equation}
which can be viewed as a transport flow in $\nu$-space. Third, we rescale $\ln S_tW$ to fix its running mean $\mu_t$ and standard deviation $\sigma_t$ to $0$ and $1$ respectively, i.e. we define $X_t\equiv(\ln S_tW-\mu_t)/\sigma_t$. The cumulants of $X_t$ are then given by $K^{(p)}_t=\psi_t^{(p)}(0)/\psi''_t(0)^{p/2}$, hence from \eqref{cgfflow}, $K^{(p)}_t=\psi^{(p)}(t)/\psi''(t)^{p/2}$. We now compute these cumulants in the $t\to\infty$ limit.

\paragraph{Positive selection (light tails).} When $p(w)$ has a light tail with index $\alpha$, Stirling's formula gives
\begin{equation}
	\psi(t)\underset{t\to\infty}{\sim} \frac{t\ln t}{\alpha} 
\end{equation}
hence for $p\geq 3$ the cumulant $K^{(p)}_t$ goes to $0$ like $t^{1-p/2}$ when $t\to\infty$. The unique distribution with vanishing cumulants is the Gaussian, hence $w$ is asymptotically log-normal with location $\mu_t\sim \ln t/\alpha$ and scale $\sigma_t\sim(\alpha t)^{-1/2}$. But since a log-normal distribution with vanishing scale is itself Gaussian, we obtain that
\begin{equation}
	S_tW\underset{t\to\infty}{\approx}\mathcal{N}\big(Ct^{1/\alpha},(\alpha t)^{-1/2}\big).  
\end{equation} 
where $\approx$ means ``is asymptotically distributed as", $\mathcal{N}(a,b)$ is a Gaussian with mean $a$ and standard deviation $b$ and $C$ is a positive constant. Note the emergence of the dynamical scaling law $t^{1/\alpha}$ for the ``speed of evolution" $\langle S_tW\rangle$ under positive selection (Fig. \ref{varianceno}).  
%
%
%
%

\paragraph{Positive selection (heavy tails).} For heavy tailed distributions we invoke Kasahara's Tauberian theorem \cite{Kasahara:1978wd,BINGHAM:1989ca} to estimate 
\begin{equation}
	\psi(\nu)\sims B'\nu^{\beta'}
\end{equation}  
where $\beta'=\beta/(\beta-1)$ is the exponent conjugate to $\beta$ and $B'$ is a positive constant which can be expressed in terms of $B$ and $\beta$. From this it follows that $\psi^{(p)}(t)$ scales like $t^{\beta'-p}$, and therefore $K^{(p)}_t$ goes to zero like $t^{\beta'(1-p/2)}$ for all $p\geq 0$. This implies that $W$ is again asymptotically log-normal as $t\to\infty$. For $\beta\leq 2$ we obtain a genuine log-normal distribution (denoted $\ln\mathcal{N}(\mu,\sigma)$ with $\mu$ the location and $\sigma$ the scale), namely
\begin{equation}
	S_tW\underset{t\to\infty}{\approx}\ln\mathcal{N}\big(B'\beta' t^{\beta'-1},\sqrt{B'\beta'(\beta'-1)\,t^{\beta'-2}}),
\end{equation}
while for $\beta> 2$ the distribution reduces to the Gaussian  
\begin{equation}
	S_tW\underset{t\to\infty}{\approx}\mathcal{N}\big(e^{B'\beta'\,t^{\beta'-1}+o(t^{\beta'-1})},\sqrt{B'\beta'(\beta'-1)\,t^{\beta'-2}}\big).
\end{equation}
In this regime the mean $\langle S_tW\rangle$ grows super-exponentially with $t$---an explosive form of selection dynamics fuelled by large amounts of initial variation. 

\paragraph{Negative selection.} When $W$ is bounded we have by Laplace's method
\begin{equation}
\psi(t)\underset{t\to\infty}{\sim}  (\ln w_+)t-\gamma\ln t
\end{equation}
 from which we compute $\lim_{t\to\infty}K^{(p)}_t=\gamma^{1-p/2} (-1)^p\,(p-1)!$. These are the cumulants of a flipped gamma distribution. Exponentiating back to $S_tW$ we obtain  
 \begin{equation}
 	 S_tW\underset{t\to\infty}{\approx}\Pi\big(\ln w_+-\gamma/t,\sqrt{\gamma}/t;\gamma\big)
 \end{equation}
 where we denoted $\Pi(\mu,\sigma;\gamma)$ the ``flipped log-gamma" distribution with density function
 \begin{equation}\label{density}
 	\pi_{\mu,\sigma}^\gamma=\frac{e^{-\mu\sqrt{\gamma}/\sigma-\gamma}\,\gamma^{\sqrt{\gamma}}}{\Gamma(\gamma)\,\sigma^{\gamma}}\, w^{\sqrt{\gamma}/\sigma-1}\left(\mu+\sigma\sqrt{\gamma}-\ln w\right)^{\gamma-1}
 \end{equation}
and support $[0,e^{\mu+\sigma\sqrt{\gamma}}]$. In the limit $\gamma\to\infty$ this gives back the log-normal distribution obtained in the previous paragraphs. That is, the continuous three-parameter family of distributions $\Pi(\mu,\sigma;\gamma)$ with $0<\gamma\leq \infty$ acts as universal attractors for the dynamics of selection. We plot the density function \eqref{density} with $\mu=0$ for several values of $\sigma$ and $\gamma$ in Fig. \ref{densityFig}. 

\paragraph{Convergence rates.} Like the location $\mu_t$ and scale $\sigma_t$, the rate of convergence of $S_tW$ to its limiting shape depends on the tail of $W$. We measure this rate by the projected relative entropy
\begin{equation}
	D(p_t\Vert \pi_{\mu_t,\sigma_t}^\gamma)=\int p_t(w)\ln\left(\frac{p_t(w)}{\pi_{\mu_t,\sigma_t}^\gamma(w)}\right)\,dw.
\end{equation}
For positive selection we find $D(p_t\Vert \pi_{\mu_t,\sigma_t}^\infty)=\mathcal{O}((K_t^{(3)})^2)$, giving a rate of convergence $\mathcal{O}(t^{-1})$ for light tails and $\mathcal{O}(t^{-\beta'})$ for heavy tails. For negative selection, assuming
\begin{equation}
	\frac{T(w)}{C(w_+-w)^{\gamma}}-1\underset{w\to w+}{\sim}Q(w_+-w)^q,
\end{equation}
for some $Q,q>0$, we compute $D(p_t\Vert \pi_{\mu_t,\sigma_t}^\gamma)=\mathcal{O}(t^{-2q})$.

\begin{figure}
	\includegraphics[width=.45\textwidth]{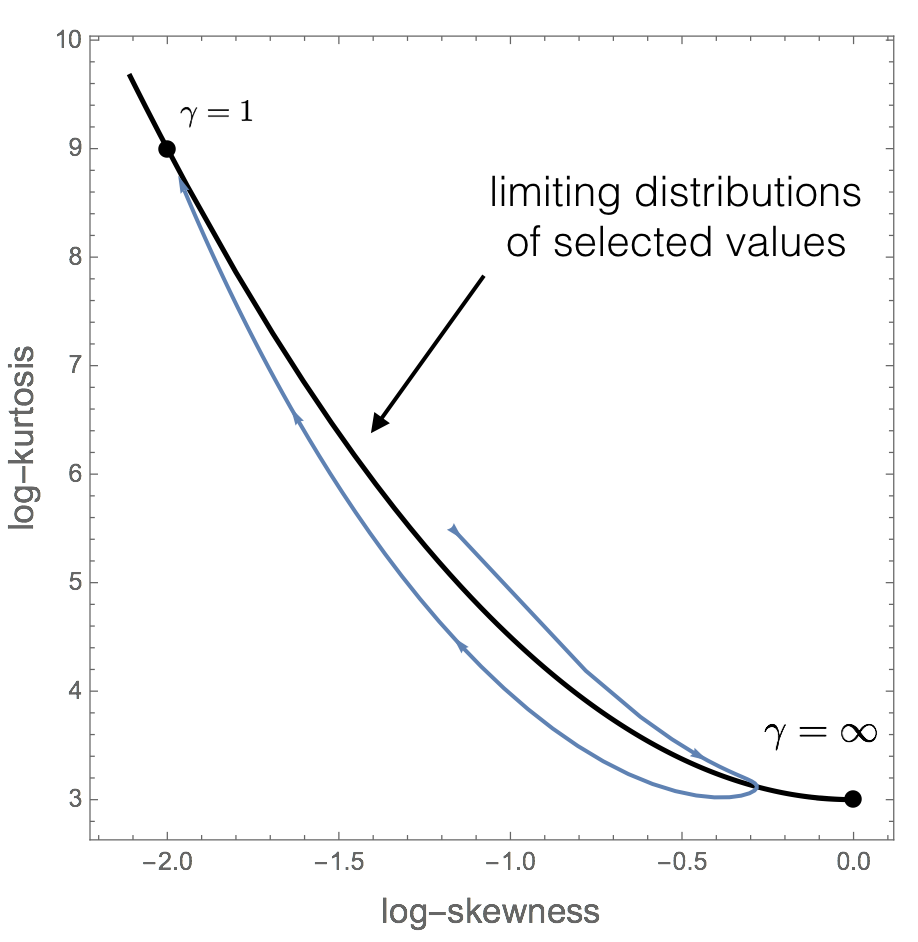}
	\caption{A truncated variable $W$ (here a unit-mean exponential truncated at $w_+=30$) is  attracted to the ``positive selection" attractor $(\gamma=\infty)$ as long as $\langle S_tW\rangle\ll w_+$, at which point it crosses over to its eventual ``negative selection" attractor ($\gamma=1)$. Here we represent this crossover in terms of the skewness and kurtosis of $\ln S_tW$ (blue line); in this plane the limiting distributions $\Pi(\mu,\sigma;\gamma)$ form a half-parabola (thick line).}
	\label{crossoverFig}
\end{figure}

\paragraph{\textbf{Crossovers and finite-size effects.}} In some cases, the evolution of $S_tW$ as $t$ increases can display a crossover between the limiting types for positive and negative selection. This arises e.g. when $W$ has a truncated distribution, such as a truncated exponential distribution $p(w)\propto\theta(w_+-w)e^{-\lambda w}$ with $\lambda w_+\gg1$. In that case $S_tW$ approches a (log-)normal distribution $\Pi(\mu,\sigma;\infty)$ as $t$ increases, until $\langle S_tW\rangle$ becomes comparable to the upper endpoint $w_+$, at which point $p_t(w)$ shifts to the negative-selection attractor $\Pi(\mu,\sigma;1)$. We can illustrate this behavior by plotting the skewness and kurtosis of $\ln S_tW$ as a function of $t$ (Fig. \ref{crossoverFig}). In this representation the universal family $\Pi(\mu,\sigma;\gamma)$ corresponds to a half-parabola where all selected values end in the limit $t\to\infty$.

It also worth emphasizing that the above results hold in the infinite population limit. For a finite population with size $N$ the scale parameter $\mu_t$ is bounded by
\begin{equation}
	\mu_+(N)\simeq p^{-1}(1/N). 
\end{equation}
For a thin-tailed distribution with index $\alpha$ this gives $\mu_+(N)=\mathcal{O}(\ln^{1/\alpha} N)$. When $\mu_t$ reaches this value, the granularity of $W$ in the tail becomes dominant, $\mu_t$ plateaus, and  our limit theorems are no longer relevant (unless a source of noise is present in the system \cite{Smerlak:2015vg}). 

\begin{figure*}
\includegraphics[width=\textwidth]{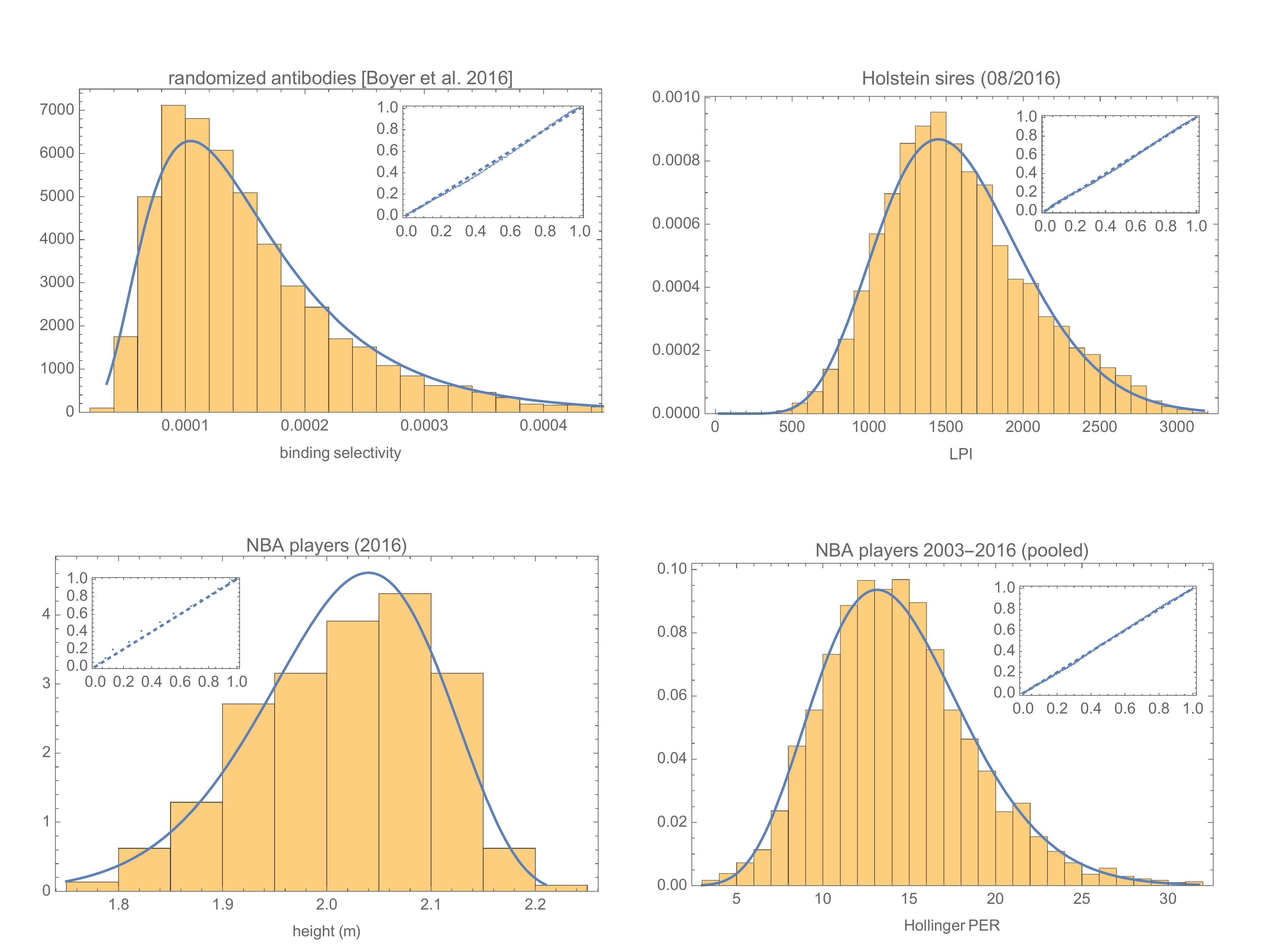}
	\caption{Empirical distributions of four natural candidates for selected values and maximum-likelihood fits by the distributions $\Pi(\mu,\sigma;\gamma)$. The insets show the corresponding probability plots. See also Table \ref{summaryStatsTab} below.}
	\label{dataFig}
\end{figure*}

\paragraph{\textbf{Datasets.}} We compared our predictions to four natural candidates for empirical selected values (Table \ref{summaryStatsTab}): the performance index (LPI) of commercial sires (selected by dairy farmers), the height and player efficiency rating (PER) of NBA players (selected by team coaches), and the selectivity of randomized antibodies with respect to a molecular target (selected by the experimental apparatus of Boyer \textit{et al.}). As shown in Fig. \ref{dataFig}, the universal family $\Pi(\mu,\sigma;\gamma)$ is a good fit to the empirical distributions, all of which are non-Gaussian ($p<10^{-18}$ or less; Pearson $\chi^2$ test). Moreover, alternative fits with the three-parameter Weibull distribution always performs worse, significantly so in three cases (Table \ref{summaryStatsTab}). We conclude that selection is a plausible explanation for the observed skewness of these variables. 

\paragraph{\textbf{Conclusion.}} We have showed that selected values have universal properties: they are attracted to a parametric family whose location, scale and shape are solely determined by the tail of the variable being selected. A parallel can be established between these results and the Fisher-Tippett-Gnedenko theorem of extreme value theory \cite{deHaan:2007bza}. Indeed, the extremality condition $M_n=\max\{X_1,\cdots X_n\}$ can also be viewed a representing an alternative form of ``selection", in which the maximum $M_n$ is picked out from the population $\{X_1,\cdots X_n\}$. This analogy is commonly made in the genetic algorithms literature, with selected and extreme values referred to as ``proportionate" and ``tournament" selection respectively \cite{Blickle:2007gd}.


Another analogy is with the Lifshitz-Slyozov-Wagner (LSW) theory of Ostwald ripening \cite{Lifshitz:1961cc,Wagner:1961gz}:\footnote{M.S. thanks Felix Otto for this analogy.} just like selected and extreme values, the size of particles in a coarsening solution follows a universal distribution characterized by the tail behavior of a suitable probability distribution. This analogy is best seen by rewriting the selection equation \eqref{sel} in terms of the density function $\pi_t(x)$, as
\begin{equation}
	\partial_t \pi_t(x)=\left(x-\int y\pi_t(y)dy \right)\pi_t(x).
\end{equation}
This integro-differential equation is similar to the LSW equation. It was showed in Ref. \cite{Niethammer:2004un} that the LSW equation has the structure of a gradient flow, and in particular has a Lyapunov functional. Whether or not a similar structure can be constructed for the selection equation is an interesting open problem.  

\begin{table*}[!t]
	\begin{minipage}{\textwidth}
  \centering
	\begin{tabular}{l*{6}{c}r}
dataset & source & selected trait $w$ & $N$ & skewness & ML $(\mu,\sigma,\gamma)$ & LLH ratio test against Weibull \\
\toprule
randomized antibodies & Ref. \cite{Boyer:2015us} & selectivity & $6,159$ & $8.36$ & $(-8.89,0.43,8.45\times 10^5)$ & $p=9.59\times 10^{-23}$\\
Holstein sires (08/2016) & CDN & LPI & $10,033$ & $0.51$ & $(7.32,0.31,27.07)$ & $p=1.32\times 10^{-28}$\\
NBA players (2016) & NBA & height ($m$) & $450$ & $-0.40$ & $(2.62,0.31,28.11)$ &$p=0.69$& \\
NBA players (2003-2016) & ESPN & Hollinger PER & $4,141$ & $0.57$ & $(0.70,0.04,7.59)$ & $p=5.66\times 10^{-9}$\\
\end{tabular}
 \bigskip
\end{minipage}
\caption{Abbreviations. NBA: National Basketball Association; CDN: Canadian Diary Network; ESPN: Entertainment Sport Programming Network; LPI: Lifetime Performance Index; PER: Player Efficiency Rating; ML: maximum likelihood; LLH: log-likelihood. The $p$-value is computed using the procedure described in Appendix C of \cite{Clauset:2009iy}.
}\label{summaryStatsTab}
\end{table*}

\section*{Acknowledgments}

Research at the Perimeter Institute is supported in part by the Government of Canada through Industry Canada and by the Province of Ontario through the Ministry of Research and Innovation.

\bibliography{library}
\end{document}